\documentclass[12pt]{article} 
\usepackage{epsfig} 
\newcommand{\beq}{\begin{equation}} 
\newcommand{\eeq}{\end{equation}} 
\newcommand{\beqs}{\begin{eqnarray}} 
\newcommand{\eeqs}{\end{eqnarray}} 
\begin{document} 
\begin{titlepage} 
\begin{flushleft} 
       \hfill                      {\tt hep-th/0011271}\\ 
       \hfill                       G\"oteborg ITP Preprint\\ 
\end{flushleft} 
\vspace*{1.8mm} 
\begin{center} 
{\bf\LARGE The Non-Abelian Self Dual String}\\ 
\vspace*{1mm} 
{\bf\LARGE on the Light Cone}\\ 
 
\vspace*{6mm} 
{\large Vanicson L. Campos\footnote{\tt vanicson@fy.chalmers.se}, 
Gabriele Ferretti\footnote{\tt ferretti@fy.chalmers.se}, 
Per Salomonson\footnote{\tt tfeps@fy.chalmers.se}}\\ 
 
\vspace*{5mm} 
{\it Institute of Theoretical  Physics\\ 
G\"{o}teborg University and Chalmers University of Technology\\ 
412 96 G\"{o}teborg, Sweden}\\ 
 
\vspace*{7mm} 
\end{center} 
\begin{abstract} 
 
\end{abstract} 
We construct the scalar profile for the non-abelian self dual 
string connecting two M5-branes compactified on a light-like circle. 
The construction is based on a conjectured modified version of 
Nahm's equations describing a  
D2-brane, with a magnetic field on it, suspended between two D4-branes. 
Turning on a constant magnetic field on the D2-brane  
corresponds to a boost in the eleventh direction. 
In the limit of infinite boost the D4-branes correspond to 
light-like compactified M5-branes. 
The solution for the scalar profile of the  
brane remains finite in this limit and displays all the correct  
expected features such as smooth interpolation between the  
unbroken and broken phase with the correct value for the Higgs  
field at infinity.  
\end{titlepage} 
 
\section{Introduction} 
 
In spite of much effort in the last five years, a satisfactory 
understanding of interacting ``non-abelian'' ${\cal N}=2$ 
superconformal\footnote{All the theories discussed in this paper
have sixteen supercharges~\cite{Seiberg1} unless specifically noted.}
field theories in six dimensions~\cite{Witten1} is still lacking. 
These theories are interesting in their own right since they correspond 
to still unknown generalizations of the concept of gauge symmetry to 
higher rank fields but also because their compactification gives rise to
a unifying description of gauge theories in various dimensions and 
with various supersymmetries. Many believe that progress in 
understanding such theories  will lead to interesting results in both 
mathematics and physics.

Given the experience of the past few years, a ``frontal attack'' on 
these theories does not look too promising and one is naturally led to
consider  limiting situations in which the physics simplifies while 
still capturing some essential features of the model. One such case is 
the breaking of the conformal symmetry by giving a vacuum expectation 
value to some of the scalars, in which case the theory becomes 
(generically) a free abelian theory in the infrared\footnote{For some 
related recent work see~\cite{Henningson, Lee}.}. 

Breaking the conformal symmetry and the non-abelian symmetry
seems a rather drastic manouvre, but there is at least one instance in 
which the broken theory retains some knowledge of the unbroken theory 
that can be extracted by semiclassical methods: the soliton sector.
There is a fairly analogous situation in four dimensions: the 
't~Hooft-Polyakov~\cite{tHooft, Polyakov} monopole versus the Dirac 
\cite{Dirac} monopole. Consider a Yang-Mills theory with gauge group 
$SU(2)$ broken to $U(1)$ by giving a non-zero v.e.v. to some scalar 
field at infinity. This theory admits smooth,  localized, finite energy,
 sourceless solutions (monopoles) where the gauge 
symmetry is restored at the center. On the contrary, a pure $U(1)$ 
theory does not admit such solutions and the monopoles in such a 
theory are singular. Studying the soliton sector of the broken theory 
thus provides important information about the theory.

Consider now the case of a six (Minkowski) dimensional theory
with ``gauge group'' $SU(2)$. It is well know that when
compactified on a small space-like circle it yields 
a five dimensional Yang-Mills theory with the same gauge
group. Such theory  has two types of solitonic excitations: in
the unbroken phase there is a point-like soliton which is nothing
other than the familiar four dimensional Euclidean instanton translated
along time, while in the broken phase there is a string-like
soliton which is just the previously mentioned 't~Hooft-Polyakov
monopole translated along the extra space direction. String
theory provides a beautiful geometrical understanding of these
two objects: First of all, the gauge theory is realized by
stacking two D4-branes on each-other \cite{Polchinski1, Witten2}
and the Higgs field describes the separation between the
branes. The monopole solution represents a D0-brane inside the
D4-brane~\cite{Witten3, Sen1} and the string-like solution is
generated by a D2-brane stretched between the two D4-branes
\cite{Strominger, Townsend}. Both such configurations preserve
half  (mutually incompatible) supersymmetries.

The full six dimensional theory can be obtained from the five
dimensional theory discussed above in the decompactification limit. 
The D4-branes become M5-branes wrapped along the non-perturbative
direction and the D2-branes are M2-branes transverse to the same
direction~\cite{Townsend}.  In this limit, the
point-like monopoles are Kaluza-Klein modes that become massless and 
are used in the formulation of the matrix model for these
theories~\cite{Aharony}, whereas the string-like solution
survives with a tension proportional to the separation between
the M5-branes in units of the eleven dimensional Planck
length. It is this last object that  will be the focus of our paper.

We will propose a method, based on a generalization of Nahm's
equations, for finding the classical profile for such ``$SU(2)$'' 
self-dual string and will present an explicit answer
for the sector of topological charge $k=1$\footnote{The 
full abelian solution was found in~\cite{Howe}. For related results
and solutions connecting a 
brane-\emph{anti}-brane pair, see~\cite{Callan,Gibbons}.}.
 
In short, the method is as  follows: Consider the perturbative  
situation in the type IIA setting with D2-branes suspended  
between D4-branes. Turning on a magnetic field on the D2-branes  
corresponds to a boost in the eleventh  
direction~\cite{Polchinski2, Chepelev}. Nahm's  
equations~\cite{Nahm}\footnote{For a clear review, see~\cite{Corrigan}.}  
are modified by the presence of such a field and so is 
the solution for the Higgs profile. In the limit of infinite boost, 
the space-like compactification radius becomes  an almost light-like  
one and we can take the limit (already considered in the matrix 
theory context~\cite{Sen2,Seiberg2}) where the light-like radius is 
kept finite in Planck units. The solution for the scalar profile of the  
brane remains finite in this limit and displays all the correct  
expected features such as smooth interpolation between the  
unbroken and broken phase with the correct value for the Higgs  
field at infinity.  
 
Given the recent frenzy of activity on strings and 
branes in the presence of electric or magnetic fields, we feel  
we should stress that our configuration is \emph{not} of the  
same type as the ones studied in~\cite{Gross, Michishita, Youm}  
and, in particular, that in our case 
the theory on the M5-brane is an ``ordinary'' one. The reason is that  
the $B$ field used here has one index tangent and  
one normal to the D4-brane and thus, can be trivially gauged away.  
 
To conclude our introduction, we should like to mention that  
our construction is based on what seems to us a very natural but 
yet not rigorously derived generalization of Nahm's construction, 
namely, the introduction of the magnetic field potential via minimal 
coupling to describe the boost of the branes. It would be interesting  
to try to provide such a derivation and to  understand better the 
solitonic structure of these fascinating theories. 
 
The paper is organized as follows: Section two  
contains a brief review of the original Nahm's construction, mainly 
to set the notation. 
In section three, we discuss the  
setup studied in this paper and the appropriate scaling limit. 
In section four we present the modified equations and solve for the 
scalar profile of the Higgs field. Section five contains the  
interpretation of the result  and the conclusions. 
 
\section{Review of Nahm's equation} 
 
Let us very briefly review Nahm's construction~\cite{Nahm} 
from the string  theory point of view~\cite{Diaconescu, Tsimpis} 
focusing, at the end, on the $k=1$ monopole sector (for recent  
reviews, see~\cite{Johnson, Nekrasov}).

The setup  consists of a pair of parallel D3-branes (yielding a 
four dimensional $SU(2)$ gauge theory after the center of mass degrees 
of freedom have been factored out) with $k$ D1-branes suspended 
between them. Let the D3-branes be placed along the $x^i$  
($i=1,2,3$) direction and let $\pm d/2$ be the position of the two  
D3-branes in the $x^4$ direction. We denote the $x^4$ direction  
by $s$: $s$ is thus 
the space coordinate on the two dimensional $SU(k)$ theory (with 
boundary) present on the stack of D1-branes. 
 
In terms of the $SU(2)$ Higgs and gauge fields $H$ and $A_\mu$ 
($\mu = 0, i$) living on the D3-branes, the BPS  
equation~\cite{Bogomolny, Prasad} for the 
static monopole solution reads\footnote{All the fields are assumed 
to be hermitian, $D_\mu H = \partial_\mu H + i[A_\mu, H]$ and 
$F_{\mu\nu} = \partial_\mu A_\nu - \partial_\nu A_\mu + 
i[A_\mu, A_\nu]$.} 
\beq 
       D_i H = \frac{1}{2}\epsilon_{ijk}F^{jk}. 
\eeq 
 
The most general charge $k$ solution to the BPS equation can be 
obtained by first considering the condition for unbroken  
supersymmetry on the stack of D1-branes, i.e. that the variation of 
the gluino field be zero. This condition can be obtained by reducing 
the same condition in ten dimensions  
$\Gamma^{MN}F_{MN} \epsilon=0$  ($M,N$ are ten dimensional indices) 
to the world volume of the D1-brane. 
In terms of the $SU(k)$ Higgs fields $T^i(s)$ 
living on the D1-branes and describing their position in relation to 
the D3-branes  this condition is precisely Nahm's equation  
\cite{Diaconescu}: 
\beq 
        \frac{d T_i}{ds} = \frac{i}{2} \epsilon_{ijk}[T^j, T^k], 
        \label{nahmfirst} 
\eeq 
to be solved with the boundary condition that the $T^i$'s have
simple poles at the positions of the D3-branes $s=\pm d/2$ with  
residues given by three matrices $t^i$ forming a $k$ dimensional  
irreducible representation of $SU(2)$.  
 
The second step is to solve the ten dimensional massless Dirac equation
$\Gamma^M D_M V=0$, again dimensionally reduced on
the D1-brane with $V = v(s)\exp(i s_i x^i)$. Here $s_i$ are the
conjugate variables to $x^i$ ($D_i = \partial/\partial s^i$) and 
$v(s)$
should be thought of as a  $2k \times 2$ matrix. The reduction  
of the Dirac equation in the presence of the D1-brane background  
yields now the associated Nahm's equation\footnote{$\sigma^i$ 
are the usual Pauli matrices.}: 
\beq 
       \left(\frac{d}{ds} -(x^i + T^i(s))\otimes \sigma^i\right) v(s)=0. 
        \label{nahmass} 
\eeq 
 
Having found a solution $v(s)$ to (\ref{nahmass}), normalized as 
\beq 
       \int_{-d/2}^{d/2} ds \, v^\dagger v = {\bf{1}}_{2\times 2},  
\label{norma} 
\eeq  
the Higgs and gauge field on the D3-brane are given respectively 
by 
\beq 
       H(x)= \int_{-d/2}^{d/2} ds \, s \, v^\dagger v \label{htp} 
\eeq 
and 
\beq 
       A_i(x) = i\int_{-d/2}^{d/2} ds \, v^\dagger  
         \frac{\partial v}{\partial  x^i}.  \label{atp} 
\eeq 
 
The $k=1$ case is particularly simple and enlightening. In this case  
the irreducible representation of $SU(2)$ is the trivial one, and  
we may set $T^i=0$. Eq. (\ref{nahmass}) with the normalization  
(\ref{norma}) is then solved  
by 
\beq 
        v= \sqrt{\frac{r}{\sinh{rd}}}e^{x_i\sigma^i s}= 
             \sqrt{\frac{r}{\sinh{rd}}}\left(\cosh(rs) + 
             \frac{x_i\sigma^i}{r} \sinh(rs) \right), \label{tp} 
\eeq 
where $r=\sqrt{x_1^2+x_2^2+x_3^2}$, and the apparent 
dimensional mismatch can be corrected by inserting the 
appropriate factors of $\alpha^\prime$ or, equivalently, by  
assigning to $d$ and $r$ opposite dimension, since they belong to  
conjugate sets of variables. It will be convenient (although slightly 
counterintuitive at first) to take $r$ to have dimension of an inverse 
length. 
Plugging (\ref{tp}) into the eqs. (\ref{htp}) and (\ref{atp}) yields 
the celebrated BPS solution for the monopole. 
 
\section{Brane Configuration and Scaling} 
\begin{figure} 
\begin{center} 
\epsfig{file=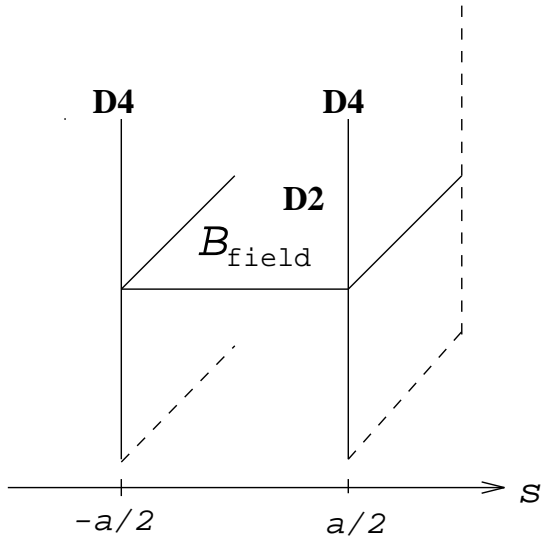,width=7 cm, angle=270} 
\caption{The brane setup used in this paper. The magnetic field $B$ 
on the D2-brane corresponds in M-theory to a boost along the  
eleventh dimension.}\label{setup} 
\end{center} 
\end{figure}

The above situation can be trivially generalized to a set  
of Dp-branes suspended between two D(p+2)-branes. What was a 
monopole in the $p=1$ case becomes now a $p-1$ dimensional 
extended object simply by trivially translating along the extra 
dimensions. In particular, our starting point (see figure~\ref{setup}) 
is the $p=2$ case  
where we have a stack of D2-branes intersecting the two  
D4-branes on a string.  
\bigskip 
\begin{center} 
\begin{tabular}{|c|c|c|c|c|c|c|c|c|c|c|} 
\hline 
&$x^0$&$x^1$&$x^2$&$x^3$&$x^4$&$x^5$&$x^6$&$x^7$&$x^8$&$x^9$\\\hline 
D4&$-$&$-$&$-$&$-$&$-d/2$&$-$&$\bullet$ 
&$\bullet$&$\bullet$&$\bullet$\\\hline 
D4&$-$&$-$&$-$&$-$&$+d/2$&$-$&$\bullet$ 
&$\bullet$&$\bullet$&$\bullet$\\\hline 
D2&$-$&$\bullet$&$\bullet$&$\bullet$&$|-|$&$-$&$ 
\bullet$&$\bullet$&$\bullet$&$\bullet$ 
\\ 
\hline 
\end{tabular} 
\end{center} 
\bigskip 

We will continue to think of $x^i$  
($i=1,2,3$) as coordinates on the D4-branes and $x^4$ as the 
coordinate along which they are separated, but we now introduce 
the coordinate $x^5$, common to all branes in the system and thus 
also to their intersection. When going to the M-theory setting we 
will denote the extra coordinate along which the M5-brane  
is wrapped by $x^\sharp$. The remaining spatial coordinates  
$x^6\cdots x^9$ play no role in our construction and are set to zero. 
 
Without further modifications, the profile for the string soliton 
on the D4-brane is exactly the same as the 't Hooft-Polyakov 
monopole. 
In this case however, we can do something extra by giving a  
D0-brane charge to the D2-brane, that is by turning on a  
constant magnetic field on the D2-brane world-volume. That  
this can be done without breaking more supersymmetries,
and that the D2-brane world volume is the only place where one
can put the D0-brane charge, is easily seen by noticing that the 
type IIA superalgebra\footnote{We use the notation  
of~\cite{Johnson} and normalize the volumes to one.} 
 
\begin{equation}  
{1\over2}  
\left\{ \left[ \begin{array}{c} Q_\alpha \\ \tilde Q_\alpha \end{array} 
\right] , \left[ Q_\beta\ \tilde Q_\beta \right] \right\} 
= \left[ \begin{array}{cc} \delta_{\alpha\beta}&0 \\  
0&\delta_{\alpha\beta} \end{array} \right] M  + 
\left[ \begin{array}{cc} 0&(\Gamma^0 Z)_{\alpha\beta} \\  
-(\Gamma^0 Z^\dagger )_{\alpha\beta}&0 \end{array} \right]\ , 
\end{equation}  
where 
\beq 
       Z=\tau_0 + \tau_2 \Gamma^{45} + \tau_4 \Gamma^{1235}, 
\eeq 
has eight zero eigenvalues (of the right chirality) for  
\beq 
       M=M_{\rm BPS}= \left|\tau_4+ 
            \sqrt{\tau_0^2 + \tau_2^2}\right|. 
\eeq 
 
It is well known that a D0-brane charge on a D2-brane  
is represented by a constant magnetic field. On the other hand,  
from the M-theory viewpoint, a D0-brane charge density  
corresponds to a momentum density along $x^\sharp$  
of the D2-brane, that is, an M2-brane moving along the  
direction $x^\sharp$. If $R$ is the compactification radius for  
$x^\sharp$, then the momentum density  
per unit area is, (neglecting numerical constants) 
\beq 
        \Pi = \frac{B}{R} = \frac{\tau_2 v}{\sqrt{1-v^2}}, 
\eeq  
where $\tau_2$ is the tension of the M2-brane, $v$  
its velocity and $B$ the magnetic field. 
 
We now study the system in a limit similar to the one of  
\cite{Sen2, Seiberg2} for matrix theory, that is, we let 
$v\to 1$ and $R/l_{Pl.}\to 0$ while keeping  
$R_- /l_{Pl.}\equiv\gamma R/l_{Pl.}$  and $d/ l_{Pl.}$ 
fixed\footnote{$l_{Pl.}$ is the M-theory Planck length that we have 
written explicitly to avoid scaling dimensionful
parameters. From now on, however, to avoid cluttering the notation,
we will always set $l_{Pl.}=1$, that is, measure
everything in M-theory Planck units.}. The system is then equivalent,  
up to an ``infinite'' boost, to a  
system of M2-branes at rest but compactified along a light-like 
circle $x^0 - x^\sharp$ of finite radius $R_-$. 
A short calculation shows that in this limit 
\beq 
        B = R_-,\qquad\hbox{in units of } l_{Pl.} 
\eeq 
that is, these two quantities can be identified when expressed in 
units of the M-theory Planck length, although, for notational 
convenience we will continue to use the quantity $B$ in solving 
Nahm's equations and revert to M-theory units only at the end. 
 
By  solving the Nahm's equations modified by the presence of such 
field and in the limit described above we will be able to derive the 
scalar profile for the $SU(2)$ self-dual string for M5-branes  
compactified on the light cone. 
 
\section{Modified Nahm's equations} 
 
The presence of a magnetic field on the D2-brane does not lead to 
any modification for the first of Nahm's equations (\ref{nahmfirst}), 
which deals with 
the center of mass degrees of freedom. The associated Nahm 
equation (\ref{nahmass}) however, is modified in a very natural 
way: consider the dimensional reduction of the ten dimensional 
Dirac equation as before but now, in order to have a non zero  
magnetic background, set $A_5 = Bs$. This choice is compatible 
with the gauge choice $A_s\equiv A_4 =0$ that had already been 
made before. 
 
We will deal explicitly only with the $k=1$ system, for which, as  
before, we can set the D2-brane Higgs fields to zero. The reduced  
Dirac equation thus becomes 
\beq 
       \left( \Gamma^4\frac{d}{ds} - i\Gamma^i x_i +  
        i Bs\Gamma^5\right) v(s)=0.  
       \label{nahm2} 
\eeq 
Notice that it is no longer possible to maintain $v$ as a  
$2\times 2$ matrix, the smallest representation for an $SO(5)$  
spinor (we need five gamma matrices) being four dimensional. 
 
It is convenient to introduce the two matrices  
\beq 
       \gamma =\frac{i x^i \Gamma^4  
       \Gamma^i}{r}\qquad\hbox{and}\qquad  
       \eta= -i \Gamma^4 \Gamma^5, 
\eeq 
satisfying $\gamma^2=\eta^2=1$ and $\{\gamma,\eta\}=0$, in terms  
of which the equation (\ref{nahm2}) becomes 
\beq 
       \left( \frac{d}{ds} - r\gamma - Bs\eta\right) v(s)=0.
       \label{nahmnew}
\eeq
We can write the most general solution, up to multiplication to the  
right by a constant matrix, as 
\beq 
       v = a(s) + b(s)\gamma + c(s)\eta + d(s)\gamma\eta, 
\eeq 
in terms of which (\ref{nahmnew}) becomes a set of coupled linear 
equations 
\beqs 
         a^\prime &=& rb+Bsc  \nonumber\\ 
         b^\prime &=& ra-Bsd  \nonumber\\ 
         c^\prime &=& rd+Bsa  \nonumber\\ 
         d^\prime &=& rc-Bsb, \label{linearset}
\eeqs 
where the prime denotes the derivative with respect to $s$.
We can scale to the dimensionless variables $x=s\sqrt{B}$ and  
$\rho^2 = r^2/B$ and define the four functions\footnote{The  
notation is chosen so that $\chi_n$ is an even (odd) function of $x$ if $n$ 
is even (odd). $\Phi(a,c,z)$ is the confluent hypergeometric  
function.} 
\beqs 
         \chi_1(\rho,x) &=&  
          \rho \, x\, \Phi(1/2 + \rho^2/4, \,  3/2, \, x^2)\, e^{-x^2/2} 
          \nonumber \\ 
        \chi_2(\rho,x)  &=& \Phi(1/2 + \rho^2/4, \, 1/2, \, x^2)\, e^{-x^2/2} 
          \nonumber \\ 
        \chi_3(\rho,x) &=& \rho \, x \, \Phi(1 + \rho^2/4, \, 3/2, \, x^2) 
              \, e^{-x^2/2}\nonumber\\ 
        \chi_4(\rho,x)  &=& \Phi(\rho^2/4, \, 1/2, \, x^2)\, e^{-x^2/2}, 
          \label{defchi} 
\eeqs 
satisfying the following relations (the prime now denotes derivative
with respect to $x$)
\beqs 
        \chi^\prime_1 + x \chi_1 = \rho \chi_2 &\quad& 
        \chi^\prime_2 - x \chi_2 = \rho \chi_1\nonumber\\ 
        \chi^\prime_3 - x \chi_3 = \rho \chi_4 &\quad& 
        \chi^\prime_4 + x \chi_4 = \rho \chi_3. 
\eeqs 
It can thus easily be checked that the most general solution to  
(\ref{nahmnew}) is given, in terms of (\ref{defchi}) and four  
integration constants $c_1\cdots c_4$ as 
\beqs 
        a &=& c_1 \chi_1 + c_2 \chi_2 + c_3 \chi_3 +  
                   c_4 \chi_4\nonumber\\ 
        b &=& c_2 \chi_1 + c_1 \chi_2 + c_4 \chi_3 +  
                   c_3\chi_4\nonumber\\ 
        c &=& -c_1 \chi_1 + c_2 \chi_2 + c_3 \chi_3 -  
                   c_4 \chi_4\nonumber\\ 
        d &=& c_2 \chi_1 - c_1 \chi_2 - c_4 \chi_3 +  
                   c_3 \chi_4.\label{gensol} 
\eeqs 
 
It is convenient to reduce (\ref{gensol}) to a subset with specific 
parity properties with respect to $x\to-x$. In analogy with the 
ordinary case we choose $a$ even ($c_1=c_3=0$) which implies 
$b$ and $d$ odd and $c$ even. With this choice, it can be easily 
 checked that  
\beqs 
   v^\dagger v &=& \bigg(2 c_2^2(\chi_1^2 + \chi_2^2) + 
   2 c_4^2(\chi_3^2 + \chi_4^2)\bigg){\bf{1}}_{4\times 4} +\nonumber\\
       &&4 c_2 c_4\bigg(\chi_2 \chi_3 + \chi_1 \chi_4\bigg)\gamma+ 
       \nonumber\\  
       && \bigg(2 c_2^2(\chi_1^2 + \chi_2^2) - 
       2 c_4^2(\chi_3^2 + \chi_4^2)\bigg) \eta. \label{vdaggerv} 
\eeqs 
Imposing the normalization condition 
\beq 
       \int_{-d/2}^{d/2} ds \, v^\dagger v = {\bf{1}}_{4\times 4}, 
\eeq 
fixes the two remaining integration constants 
\beqs 
       c_2 &=& \frac{B^{1/4}}{2} 
         \left(I_1(\delta, \rho) \right)^{-1/2}\nonumber \\ 
        c_4 &=& \frac{B^{1/4}}{2} 
         \left(I_2(\delta, \rho) \right)^{-1/2},\label{i1i2} 
\eeqs 
where we defined one more dimensionless constant  
$\delta=d\sqrt{B}$ and the two integrals: 
\beqs 
       I_1(\delta, \rho)  &=& \int_{-\delta/2}^{\delta/2} dx 
                 (\chi_1^2(\rho,x)  + \chi_2^2(\rho,x) ) \nonumber \\ 
       I_2(\delta, \rho)  &=&  \int_{-\delta/2}^{\delta/2} dx 
                 (\chi_3^2(\rho,x) + \chi_4^2(\rho,x) ).  
\eeqs 
 
When lifting to
M-theory it is convenient to think of $R_-$ instead of $B$.  
The Higgs field on the light-like compactified
M5-branes is still given by (\ref{htp}) and becomes,  
(considering only its length) 
\beq
|H(r)| = \frac{1}{\sqrt{R_-}}\frac{K(d\sqrt{R_-}, r/\sqrt{R_-})}
                   {\left(I_1(d\sqrt{R_-}, r/\sqrt{R_-})\,\, 
                          I_2(d\sqrt{R_-}, r/\sqrt{R_-})\right)^{1/2}},
       \label{final}
\eeq
with the integral $K$ being defined as
\beq 
       K(\delta, \rho)  = \int_{-\delta/2}^{\delta/2} dx \, 
                 x \, (\chi_2(\rho,x) \chi_3(\rho,x)  +  
                 \chi_1(\rho,x) \chi_4(\rho,x) ). 
\eeq 
\begin{figure}
\begin{center}
\epsfig{file=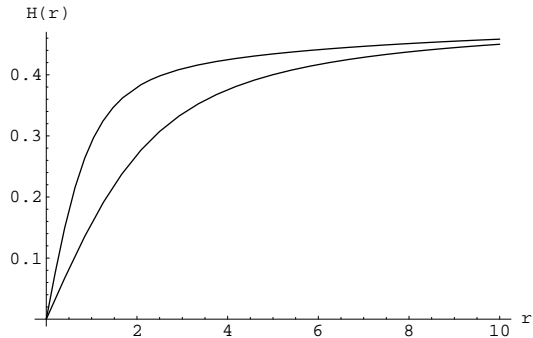,width=7 cm}
\caption{Dependence of the scalar field $H$ on the distance $r$
for two values of the light-like compactification radius.
The steeper curve corresponds to $R_-=4$ and the flatter to
$R_-=1$.}\label{higgsy}
\end{center}
\end{figure}

\section{Conclusions} 
 
Equation (\ref{final}) is our result for the profile of the brane.  
The Higgs field has an implicit dependence on the asymptotic  
separation of the branes at infinity $d$ and on the light-like 
compactification radius $R_-$.  
Let us analyze this result in the M-theory context.  
A plot of $H(r)$ for $d=1$ and for two different values of $R_-$ is  
given in figure~\ref{higgsy}. 
 
Various simple consistency checks can be performed.  
First of all, it can be noticed that the asymptotic 
value is independent on $R_-$ as it should. Also, the curve
goes to zero at the origin yielding a smooth solution  
everywhere. The solution becomes steeper and steeper as we 
let $R_- \to \infty$. This is also expected. In this limit,  
our solution is just one of an infinite number of modes~\cite{Lee} 
becoming massless as 
\beq 
       H^{(l)} \approx \left( \delta_{l,0} -  
        \frac{1}{2R_- r}e^{-l r /R_-}\right). 
\eeq 
 
These characteristics give us confidence that we have found a way 
of describing these solitonic excitations. It would be interesting 
to test this further and to see whether it is possible to use these 
techniques to describe the full (uncompactified) six  
dimensional solution, including the three-form field. 
 
\section{Acknowledgments} 
We wish thank M{\aa}ns Henningson and Dimitrios Tsimpis for discussion.

\end{document}